\documentclass[twocolumn,superscriptaddress,
 amsmath,amssymb,
 aps,
 prl,
]{revtex4-2}
\usepackage{dcolumn}% Align table columns on decimal point
\usepackage{bm}% bold math
\usepackage{graphicx}
\usepackage{comment}
\usepackage[caption=false]{subfig}
\usepackage{color}

\begin{document}

%\preprint{}
\title{Optimal Resetting Brownian Bridges}% Force line breaks with \\
\author{Benjamin De Bruyne}
\affiliation{LPTMS, CNRS, Univ.  Paris-Sud,  Universit\'e Paris-Saclay,  91405 Orsay,  France}
\author{Satya N. Majumdar}
\affiliation{LPTMS, CNRS, Univ.  Paris-Sud,  Universit\'e Paris-Saclay,  91405 Orsay,  France}
\author{Gr\'egory Schehr}
\affiliation{Sorbonne Universit\'e, Laboratoire de Physique Th\'eorique et Hautes Energies, CNRS UMR 7589, 4 Place Jussieu, 75252 Paris Cedex 05, France}

%\date{\today}

\begin{abstract}
We introduce a resetting Brownian bridge as a simple model to study search processes where the total search time $t_f$ is finite and the searcher returns to its starting point at $t_f$. This is simply a Brownian motion with a Poissonian resetting rate $r$ to the origin which is constrained to start and end at the origin at time $t_f$. We first provide a rejection-free algorithm to generate such resetting  bridges in all dimensions by deriving an effective Langevin equation with an explicit space-time dependent drift $\tilde \mu({\bf x},t)$ and resetting rate $\tilde r({\bf x}, t)$. We also study the efficiency of the search process in one-dimension by computing exactly various observables such as the mean-square displacement, the hitting probability of a fixed target and the expected maximum. Surprisingly, we find that there exists an optimal resetting rate $r^*$ that maximizes the search efficiency, even in the presence of a bridge constraint. We show however that the physical mechanism responsible for this optimal resetting rate for bridges is entirely different from resetting Brownian motions without the bridge constraint. 
\end{abstract}

\maketitle

%\tableofcontents
\newpage

Search processes are ubiquitous in nature. They appear in a wide range of situations ranging from foraging animals \cite{Bartumeus09,Viswanathan11}, biochemical reactions \cite{Berg81,Coppey04,Ghosh18,Chowdhury19} and all the way to behavioral psychology \cite{Wolfe04,Bell91,Adam68}. Search problems exhibit rich features \cite{Montanari02,Gelenbe10,Snider12,Abdelrahman13,Chupeau17,Oshanin07} and finding an optimal search strategy in a given context is an interesting problem with multiple applications
across disciplines~\cite{Benichou11,Lomholt08}. In recent years, there has been a surge of interest in the effect of resetting in search processes (for a recent review see \cite{Evans20}). Stopping and starting from scratch has shown to be an efficient search strategy in several contexts such as in optimization algorithms \cite{Villen91,Luby93,Tong08,Avrachenkov13,Lorenz18}, chemical reactions \cite{Reuveni14,Reuveni15}, animal foraging \cite{Boyer14,Majumdar15b,Boyer18,Maso19c,Pal20} and catastrophes in population dynamics \cite{Levikson77,Pakes78,Pakes97,Brockwell82,Brockwell85,Kyriakidis94,Manrubia99,Economou03,Visco10,Dharmaraja15}.
Perhaps, the effect of resetting is best seen in the simple model of diffusion introduced by Evans and Majumdar~\cite{Evans11}. 
In this resetting Brownian motion (RBM) model, the position $x(t)$ of a Brownian motion, e.g. in one dimension, is reset to the origin randomly in time according to a Poisson process with a constant rate $r$. In a time interval $dt$, the position $x(t)$ follows the stochastic rule
\begin{align}
  \hspace{-0.5em} x(t\!+\!dt)\!=\!\left\{\begin{array}{ll}
    \hspace{-0.4em}x(t)\!+\!\sqrt{2D}\,\eta(t)dt, &\text{with \, prob. } 1\!-\!r\,dt,\\
    \hspace{-0.4em}0,&\text{with \, prob. } \!r\,dt,
  \end{array}\right.\label{eq:eom}
\end{align}
where $D$ is the diffusion coefficient and $\eta(t)$ is an uncorrelated white noise with zero mean $\langle \eta(t)\rangle=0$ and delta correlator $\langle\eta(t)\eta(t')\rangle=\delta(t-t')$. The dynamics therefore consists in a combination of pure diffusion with intermittent resets to the origin. The effect of resetting on the search process can be simply measured by the mean first-passage time $\langle T(M)\rangle$ to a level $M$, which is the mean time the searcher takes to find a target located at a position $M$.  For pure diffusion without resetting, it is well-known that this quantity is infinite \cite{RednerGuide,Bray13}. In contrast, resetting leads to the striking result that the mean first-passage time $\langle T(M)\rangle$ becomes not only finite but also that it becomes minimal at
an optimal resetting rate $r^*$. The mechanism behind this result is that resetting suppresses the trajectories that diffuse far away from the target and makes them restart from the origin, hence increasing their chances to find the target. Since the original model, the existence of an optimal resetting rate has been studied extensively for various stochastic processes, leading to a tremendous amount of activities \cite{Evans11b,Montero16,Evans14,Christou15,Evans18,Nagar16,Pal16,Eule16,Kusmierz14,Campos15,Bressloff20a,Bressloff20b,Reuveni16,Pal17,Sokolov18,Prasad19a,Prasad19b,DeBruyne20a,DeBruyne20b,DeBruyneMori21} -- see \cite{Evans20} for a review.

The existence of this optimal resetting rate has also been confirmed in experiments with optical traps in both one and two dimensions~\cite{Tal20,Besga20,Faisant21}. 

In most examples of search processes with resetting, the underlying stochastic process is {\it free}, in the sense that it does not satisfy any additional constraints. However, in most
circumstances, search processes are typically time-limited and do not continue for ever. For example, in the context of animals foraging for food \cite{Boyer14, Boyer18, Pal20,Shlesinger06,Giuggioli05,Randon09,MajumdarCom10,Murphy92,Boyle09}, they typically start from
their nest and come back to it at the end of the day. Similarly, in a search and rescue mission after a plane crash in the sea, the divers are typically called
off after a certain fixed amount of time and they go back to their initial location. Here, the underlying search processes are {\it constrained} to come back to their
starting point after a fixed time $t_f$. For example, in the context of random Brownian search, the trajectories would form a stochastic Brownian bridge of duration $t_f$. 
Thus a natural question arises: when the search time $t_f$ is fixed and the searcher is constrained to go back to the initial position at $t_f$, is resetting still a good search strategy? 
Moreover, does the paradigm of an optimal resetting rate $r^*$ still exist in the case of a stochastic bridge? Another natural aspect of this time constrained search processes is an algorithmic one: how does one generate such stochastic bridges with resetting with the correct statistical weight in an efficient rejection-free manner? In this Letter, we address these two important questions.

\begin{figure}[t]
   \subfloat{%
 \includegraphics[width=0.24\textwidth]{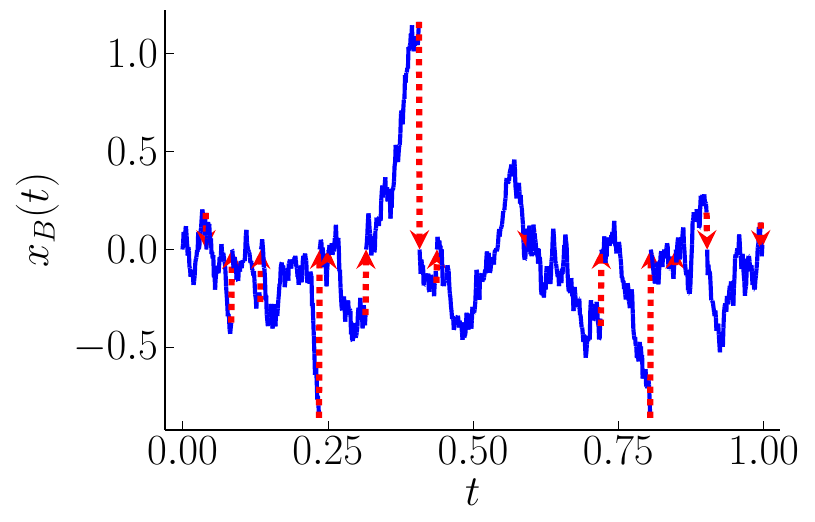}%
}\hfill
\subfloat{%
  \includegraphics[width=0.24\textwidth]{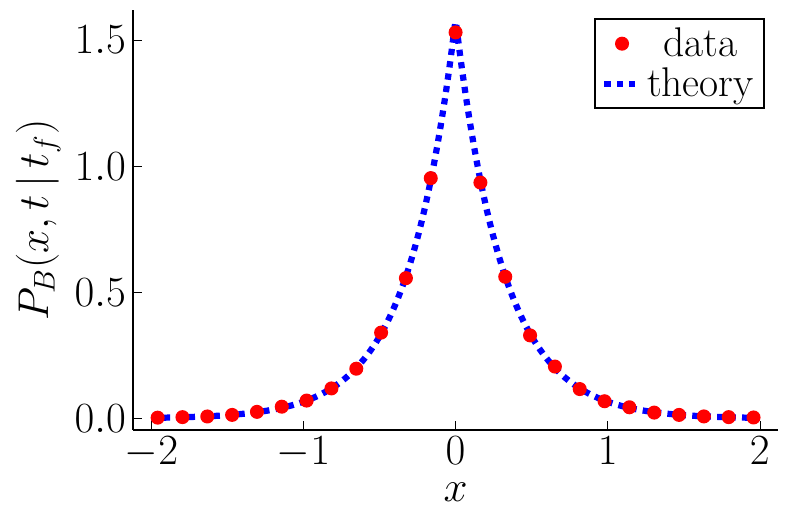}%
}
    \caption{\textbf{Left panel:} A typical RBB trajectory $x_B(t)$ with resetting rate $r=10$, diffusion constant $D=1$ and duration $t_f=1$. The resetting events are denoted by red dashed lines with arrows. \textbf{Right panel:} Position distribution for an RBB at an intermediate time $t=t_f/2$ where $r=10$, $D=1$ and $t_f=1$. The distribution obtained numerically by sampling the trajectories from the effective Langevin equation in Eq.~(\ref{eq:eomeff}) is compared with the theoretical prediction in Eq.~(\ref{eq:Pb}) -- see Eq.~(17) in \cite{supp} for a more explicit expression. }
    \label{fig:B}
\end{figure}

We start with a one-dimensional resetting Brownian \emph{bridge} (RBB) -- see the left panel in Fig.~\ref{fig:B} -- whose position $x_B(t)$ evolves locally according to the stochastic rule in Eq.~(\ref{eq:eom}) and is constrained to return to the origin at a future time $t_f$, i.e.~$x_B(0) = x_B(t_f) = 0$. We first propose a completely rejection-free algorithm for generating an RBB -- we show that this can be obtained by imposing a drift towards the origin and an effective resetting rate and that both are space-time dependent [see Eqs.~(\ref{eq:eomeff})-(\ref{eq:VW})]. Next, we investigate the mean-square fluctuations of the RBB. In the presence of a finite resetting rate $r$ to the origin, one would have naively expected that any finite resetting, in addition to the bridge constraint $x_B(t_f) = 0$, will localize the trajectories towards the origin and hence the mean-square fluctuations will decrease monotonically with increasing $r$. Surprisingly, we find that, as a function of $r$, the mean-square fluctuations is non-monotonic -- it first
increases with $r$, achieves a maximum at some $r^*$ and then decreases monotonically for increasing $r$. Thus there is indeed an optimal $r^*$ that maximizes the spatial fluctuations. We show that there is a rather interesting physical mechanism involving a trade-off between the bridge constraint and the resetting that leads to this optimal resetting rate. This has important consequences for search processes. If a target is placed at a fixed distance from the origin where the searcher starts, resets and returns, a bigger spatial fluctuation of the searcher would imply that the target will be found more easily. Indeed, we also compute exactly the probability to find the target before $t_f$ (hitting probability) and show that it also has a non-monotonic behavior as a function of $r$, achieving its maximum at an optimal resetting rate.  Computations of other observables, such as the expected maximum of the RBB, also confirms the existence of an optimal resetting rate. Thus the paradigm of the existence of an optimal resetting rate also holds for constrained bridge processes with resetting, albeit the physical mechanism at play is rather different from the free case.

Our first goal is to construct a rejection-free algorithm to generate an RBB with the correct statistical weight. Generating constrained stochastic Markov processes was initially studied in the probability literature \cite{Doob,Pitman} and more recently it has emerged as a vibrant research area by itself in the context of sampling rare/constrained trajectories with 
applications ranging from chemistry and biology all the way to particle physics~\cite{BCDG2002,GKP2006,GKLT2011,KGGW2018,Gar2018,Rose21,Rose21area,CLV21,CT2013,MajumdarEff15,Orland,Grela2021,DebruyneAR21,Mazzolo17a,Mazzolo17b,Monthus21,Baldassarri21,DebruyneRW21,DebruyneRTP21,Brunet20}. In the context of the RBB, a naive solution 
would be to generate free RBM paths and discard the ones that do not satisfy the bridge constraint $x_B(t_f) = 0$. However, this is computationally wasteful since the RBM trajectories that go back to the vicinity of $0$ at $t=t_f$ are typically rare. Here, we show that the RBB trajectories can be generated in a rejection-free manner, from the effective Langevin equation
 \begin{eqnarray}
  x_B(t\!+\!d t)\! =\!\left\{\begin{array}{ll}
    x_B(t)\!+\!\sqrt{2D}\,\eta\, d t\!+\!\tilde \mu dt, &p=1\!-\!\tilde rdt,\\
    0,&p=\tilde r\,dt,
  \end{array}\right.\label{eq:eomeff}
\end{eqnarray}
where the effective drift $\tilde \mu$ and resetting rate $\tilde r$ are space-time dependent and are given by
\begin{subequations}
\begin{align}
\tilde \mu(x_B,t) &= -\sqrt{4rD} \,\mathcal{V}\left(y,\tau\right),\label{eq:effmus}\\
  \tilde r(x_B,t) &= r\, \mathcal{W}\left(y,\tau\right),\label{eq:effrs}
\end{align}
\label{eq:mur}
\end{subequations}
\hspace*{-0.1cm}with $y=x_B/\sqrt{4D(t_f-t)}$, $\tau=r(t_f-t)$ and the scaling functions 
\begin{subequations}
\begin{align}
\mathcal{V}(y,\tau) &= \frac{y\, e^{-y^2 -\tau}}{\sqrt{\tau}\left[e^{-\tau-y^2}+\sqrt{\pi \tau}\,\text{erf}(\sqrt{\tau})\right]}\label{eq:V}\,,\\
  \mathcal{W}(y,\tau) &= \frac{e^{-\tau} + \sqrt{\pi \tau}\,\text{erf}(\sqrt{\tau})}{e^{-\tau-y^2}+\sqrt{\pi \tau}\,\text{erf}(\sqrt{\tau})}\,.\label{eq:W} 
\end{align}
\label{eq:VW}
\end{subequations}
\hspace*{-0.2cm}Here ${\rm erf}(z) = (2/\sqrt{\pi}) \int_{0}^z e^{-u^2}\, du$ is the error function. Note that while RBM with space-time dependent resetting rates have been studied before \cite{RG2017,Shk17,Pinsky20,Evans20,Evans11b,Ray20,BS20}, here they emerge naturally and
have a specific form in order to satisfy the bridge constraint. 

To derive this effective Langevin equation (\ref{eq:eomeff}), we consider the probability distribution function (PDF) $P_B(x,t\,|\,t_f)$ of the position $x_B(t)$
of an RBB of total duration $t_f$. We split the interval $[0,t_f]$ into two parts: $[0,t]$ and $[t,t_f]$ and use the Markov property of the bridge to write
\begin{align}
  P_B(x,t\,|\,t_f) =\frac{P_r(x,t\,|\,0,0)P_r(0,t_f\,|\,x,t)}{P_r(0,t\,|\,0,0)} \;,\label{eq:Pb}
\end{align}

where $P_r(x,t\,|\,0,0)$ is the PDF of the RBM at time $t$, starting from the origin at $t=0$. The denominator is just a normalization constant that ``counts'' all the trajectories of the RBM of duration $t_f$, starting and ending at $0$. Note that Eq.~(\ref{eq:Pb}) can be interpreted as the fraction of all RBM paths of duration $t_f$ satisfying the bridge constraint that also pass through $x$ at time $t$. To ease notation, we introduce the \emph{forward} propagator $P_r(x,t)\equiv P_r(x,t\,|\,0,0)$, where $x$ appears as the final position, and the \emph{backward} propagator $Q_r(x,t)\equiv P_r(0,t_f\,|\,x,t)$, where $x$ appears as the initial position. In these notations, we have that $P_B(x,t\,|\,t_f)=P_r(x,t)Q_r(x,t)/P_r(0,t_f)$. It is well-known that $P_r(x,t)$ and $Q_r(x,t)$ satisfy the forward and backward Fokker-Plank equations of RBM respectively given by (see Supp. Mat. \cite{supp})
\begin{subequations}
\begin{align}
  \partial_t P_r(x,t) &\!=\! D \partial_{xx} P_r(x,t)\!-\!r P_r(x,t)\!+\!r \delta(x),\label{eq:fP}\\
   -\partial_t Q_r(x,t) &\!=\! D \partial_{xx} Q_r(x,t)\!-\!r Q_r(x,t)\!+\!r Q_r(0,t),\label{eq:bP}
\end{align}
\label{eq:fbP}
\end{subequations}
\hspace*{-0.2cm}with the initial and final conditions $P_r(x,0)=\delta(x)$, $Q_r(x,t_f)=\delta(x)$. Our goal is to write the Fokker-Plank equation satisfied by the bridge PDF $P_B(x,t\,|\,t_f) $. Taking a time derivative of Eq.~(\ref{eq:Pb}) and using Eqs.~(\ref{eq:fbP}) satisfied by the free propagators, we get \cite{supp}
\begin{align}
  & \partial_t P_B(x,t\,|\,t_f)\!=\!D \,\partial_{xx} P_B(x,t\,|\,t_f)\!-\!\partial_x\left[\tilde \mu(x,t) P_B(x,t\,|\,t_f)\right] \nonumber\\
   &\quad\!-\!\tilde r(x,t) P_B(x,t\,|\,t_f)\!+\!\int_{-\infty}^\infty \tilde r(x',t) P_B(x',t)dx'\,,\label{eq:effP2}
\end{align}
where we have introduced an effective space-time dependent drift $\tilde \mu(x,t)$ and resetting rate $\tilde r(x,t)$ which are given~by
\begin{equation}
 \tilde \mu(x,t) = 2D\,\partial_x \ln(Q_r(x,t)) \;;\;   \tilde r(x,t) = r\, \frac{Q_r(0,t)}{Q_r(x,t)}  \;.
\label{eq:effmur}  
\end{equation}
One can show that the effective Langevin equation corresponding to the Fokker-Planck equation in (\ref{eq:effP2}) is given exactly by Eq.~(\ref{eq:eomeff}). To compute $ \tilde \mu(x,t) $ and $ \tilde r(x,t) $, we need to compute the backward propagator $Q_r(x,t)$ in Eq.~(\ref{eq:effmur}). Noting that $Q_r(x,t) = P_r(0,t_f|x,t)$, we just need the propagator for the RBM, which can be computed by using the renewal identity~\cite{Evans20}
\begin{align}
  P_r(x,t\,|\,x_0,0) &= e^{-rt}P_0(x,t\,|\,x_0,0) \nonumber\\
  &\quad + r \int_0^t d\tau e^{-r\tau}P_0(x,\tau\,|\,0,0)\,,\label{eq:ren}
\end{align}
where $P_0(x,t\,|\,x_0,0)=e^{-\frac{(x-x_0)^2}{4Dt}}/\sqrt{4\pi Dt}$ is the standard Brownian propagator (without resetting). The renewal identity (\ref{eq:ren}) simply states that for the particle to be at $x$ at a time $t$, it either (i) must never reset, in which case its probability distribution is just the one of a free Brownian motion $P_0(x,t\,|\,x_0,0)$, or (ii), reset for the last time at $t-\tau>0$, after which the particle restarts from the origin and then propagates to $x$ in the remaining time $\tau$. As the resetting times follow a Poisson process, the former event happens with probability $e^{-rt}$ while the latter happens with probability $re^{-r\tau}$ and has to be summed over all $\tau$ in $[0,t]$. From the renewal identity, one can straightforwardly obtain $Q_r(x,t)$ and then, using (\ref{eq:effmur}), find the exact expressions for $\tilde \mu(x,t)$ and $\tilde r(x,t)$ as given in Eqs.~(\ref{eq:mur}) and (\ref{eq:VW}). The effective Langevin equation in Eq.~(\ref{eq:eomeff}) can then be used to generate RBB trajectories (see left panel in Fig.~\ref{fig:B}). Furthermore, the distribution of the position $P_B(x,t\,|\,t_f)$ obtained numerically is in excellent agreement with the theoretical one obtained in Eq.~(\ref{eq:Pb}) (see right panel in Fig.~\ref{fig:B}). The effective Langevin equation (\ref{eq:eomeff}) derived for the one-dimensional RBB can be generalized to RBB in higher dimensions in a rather straightforward manner, as detailed in~\cite{supp}.

\begin{figure}[t]
   \subfloat{%
 \includegraphics[width=0.24\textwidth]{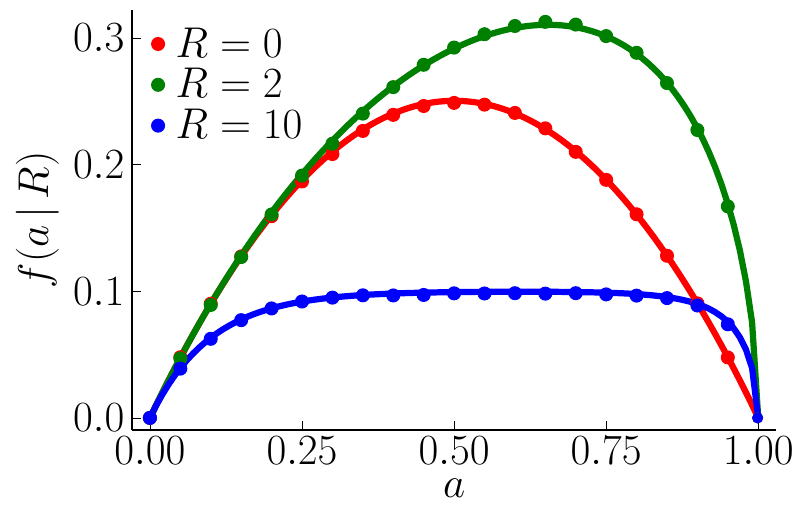}%
}\hfill
\subfloat{%
  \includegraphics[width=0.24\textwidth]{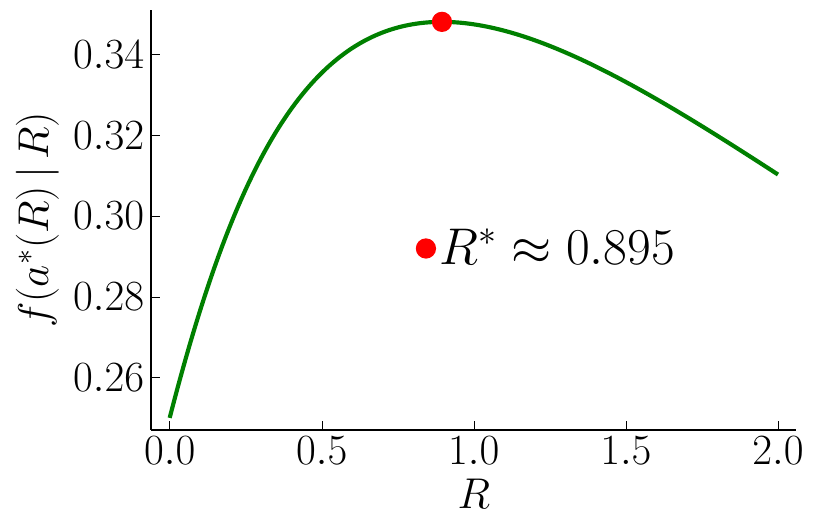}%
}
\caption{\textbf{Left panel:} The function $f(a|R)$ plotted vs.~$a$ is evaluated numerically using the effective Langevin equation in Eq.~(\ref{eq:eomeff}) (symbols) and is compared to the theoretical prediction (plain lines), given in Eq.~(\ref{msd_rbb.2}), for different values of $R$ -- see also Eqs.~(22) and (24) in \cite{supp}.
This function is clearly asymmetric around $a=1/2$. Only when $R\to 0$,
it approaches to the symmetric form $f(a|R\to 0)=a(1-a)$. For any $R$, the function $f(a|R)$ has a unique maximum at $a=a^*(R)$. \textbf{Right panel:}  The maximal value $f(a^*(R)|R)$ plotted vs. $R$.
It has a unique maximum at $R^*\approx 0.895$ (red dot).}
\label{fig.far_1}
\end{figure}

Let us now consider the RBB as a search process and show that resetting enhances its search efficiency through a mechanism that is quite different from the one underlying the RBM. Below we illustrate this enhancement by studying three different measures: the mean square displacement, the hitting probability and the expected maximum of this process.
\paragraph{Mean square displacement.} The PDF of the position $x_B(t)$ of an RBB at some intermediate time $0\le t \le t_f$
is given in Eq.~(\ref{eq:Pb}). The mean position $\langle x_B\rangle (t\,|\,t_f)$ vanishes by symmetry. Hence the minimal quantity that characterizes the spatial fluctuations is the second moment of the PDF, i.e., the mean-square displacement $\langle x_B^2\rangle (t|t_f)$. We compute $\langle x_B^2\rangle (t|t_f)$ from Eq.~(\ref{eq:Pb}) explicitly in \cite{supp}, leading to
\begin{equation}
\langle x_B^2\rangle (t\,|\,t_f)= 2D\, t_f \, f\left(a=\frac{t}{t_f}\,\bigg|\,R= r\, t_f\right) ,
\label{msd_rbb.2}
\end{equation}
where the scaling function $f$ is given in \cite{supp}. 
A plot of the function $f(a|R)$ vs. $a\in [0,1]$,
for different values of $R$, is given in the left panel in Fig.~\ref{fig.far_1}. 
As the rescaled resetting rate $R= r \,t_f$ varies from $0$ to $\infty$, the function $f(a|R)$, crosses over from a 
parabolic to a flat shape, i.e., $f(a|R\to 0)  = a(1-a)$ and $f(a|R\to \infty) \approx 1/R$. 
For a general $R$, the function $f(a|R)$ is not symmetric around $a=1/2$,
since resetting breaks the time-reversal symmetry. 
For a given $R$, the function $f(a|R)$ has a unique maximum at $a=a^*(R)$ and
this maximal mean square displacement
$f(a^*(R)|R)$ (in units of $2 Dt_f$), as a function of $R$, has a non-monotonic behavior:
it first increases with increasing $R$, achieves a maximum at
$R=R^*\approx 0.895$ and then decreases again with increasing $R$ (see Fig.~\ref{fig.far_1}). Thus, interestingly, a nonzero resetting rate, when it is not too large, actually enhances the bridge fluctuations. Naively, one would think
that resetting to the origin localizes the trajectory of the bridge in the vicinity
of $x=0$ and thus would suppress fluctuations. This naive picture holds only for very large $R$. Moreover, there is a non-trivial optimal
rescaled resetting rate $R^*$ that optimizes the maximum value of the mean-square displacement $f(a^*(R)|R)$ over the full interval $[0,t_f]$, thus enabling
the particle to explore more space. The physical mechanism behind this surprising result can be understood as follows. In the absence of resetting, 
the particle cannot go too far away from the origin, since it has to come back to the origin at time $t_f$, by a slow diffusing process. However, when a small amount of
resetting rate $r$ is switched on, the particle can go further away from the origin since it can come back to the origin at time $t=t_f$ by a ``last minute'' instantaneous resetting. 
Hence there is a subtle trade-off between the resetting and the bridge constraint. This mechanism for an optimal $r^*$ in the RBB is thus very different from the one in the free RBM.

\begin{figure}[t]
\subfloat{%
 \includegraphics[width=0.24\textwidth]{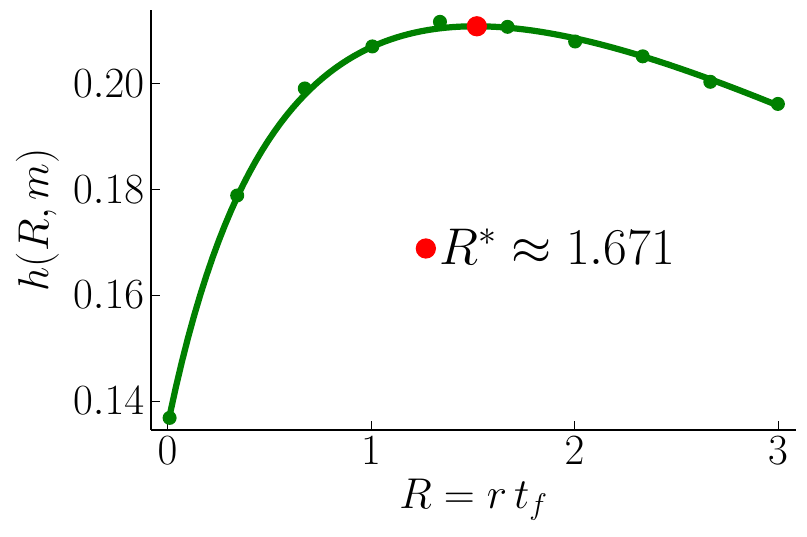}%
}\hfill
\subfloat{%
  \includegraphics[width=0.24\textwidth]{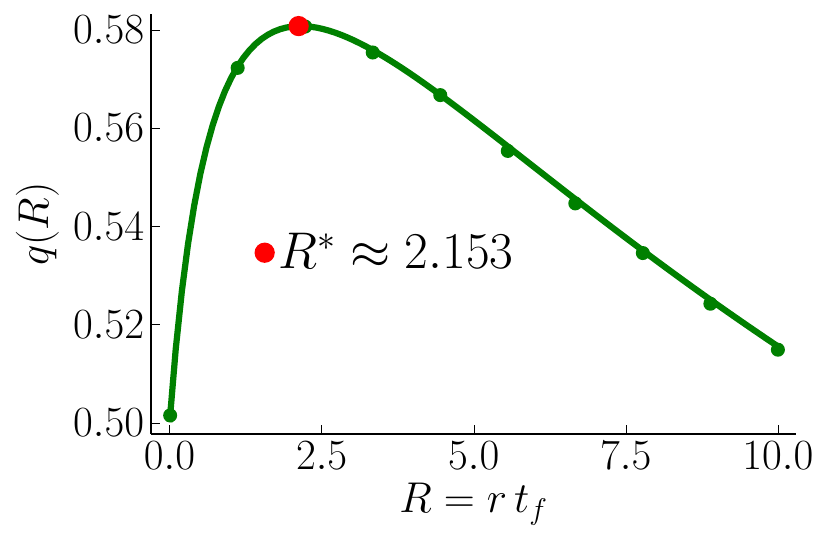}%
}
\caption{\textbf{Left panel:} The theoretical prediction (solid line) of the hitting probability in Eq.~(\ref{eq:phits}) with $m=1$ (more explicit form given in Eq.~(39) in \cite{supp}), compared with the one evaluated numerically from the effective Langevin equation~(\ref{eq:eomeff}) (symbols) with $D=1$ and $t_f=1$. For a given $m$, it exhibits a unique maximum at $R=R^* \approx 1.671$ (red dot). \textbf{Right panel:} The theoretical prediction (solid line) of the rescaled expected maximum $\langle M(t_f)\rangle=\sqrt{\pi D\,t_f}\, q(R=r t_f)$ in Eq.~(52) in \cite{supp} compared with the one evaluated numerically (symbols) with $D=1$ and $t_f=1$. The function $q(R)$ has a maximum at $R^*\approx 2.153$. }
\label{fig:probh}
\end{figure}

\paragraph{Hitting probability.}  To further explore this trade-off mechanism between the resetting and the bridge constraint in the context of a search for a fixed target located at $M$, we next compute the hitting probability, i.e., the probability that the RBB (searcher) finds the target at $M$ before time $t_f$. The hitting probability can be computed from the relation
\begin{align}
  p_{\text{hit}}(t_f,M) = \int_0^{t_f} dt F_B(t\,|\,M,t_f)\,,\label{eq:phit}
\end{align}
where $F_B(t\,|\,M,t_f)$ is the first-passage probability density of the RBB at level $M$ with $t \leq t_f$. This can be computed by decomposing the RBB trajectories into two parts: one in the time interval $[0,t]$ where it first hits the level $M$ at a time $t<t_f$, another one in the time interval $[t,t_f]$ where it propagates from $M$ to the origin. One gets
\begin{align}
  F_B(t\,|\,M,t_f) = \frac{F_r(t\,|\,M) P_r(0,t_f\,|\,M,t)}{P_r(0,t_f\,|\,0,0)}\,,\label{eq:FB}
\end{align}
where $F_r(t\,|\,M)$ is the first-passage time distribution of a RBM \cite{Evans11}, $P_r(x,t\,|\,x_0,t)$ is the propagator of a RBM given in Eq.~(\ref{eq:ren}) and the denominator is a normalization factor that ``counts'' all the bridge trajectories. Using the known results for $F_r(t\,|\,M)$ \cite{Evans11} and the propagator from Eq.~(\ref{eq:ren}) (see \cite{supp} for details) we get
\begin{align}
   p_{\text{hit}}(t_f,M) &= h\left(R=rt_f,m=\frac{M}{\sqrt{2D t_f}}\right)\,,\label{eq:phits}
\end{align}
where the scaling function $h$ can be represented as a Bromwich integral (see \cite{supp}) 
\begin{align}
  h(R,m) \!=\!c(R)\!\int_\Gamma \frac{du\,e^{u}}{2\pi i}  \frac{\sqrt{u\!+\!R}}{u}\,\frac{R\!+\!u\, e^{-m\sqrt{2}\sqrt{u+R}}}{R\!+\!u\, e^{m\sqrt{2}\sqrt{u+R}}}\,\label{eq:hRy} \;,
\end{align}
with $c(R)= \sqrt{\pi }/[\sqrt{\pi R}\,\text{erf}\left(\sqrt{R}\right)+e^{-R}]$. When $R=0$, we recover the hitting probability of a Brownian bridge $h(R=0,m)=e^{-2m^2}$. For a given target position $m$, the function $h(R,m)$, as a function of $R$, has a non-monotonic behavior and achieves a maximum at $R=R^*(m)$. A plot of $h(R,m)$ vs $R$ for $m=1$ is shown in the left panel in Fig.~\ref{fig:probh}. Thus the paradigm of an optimal resetting rate $R^*(m)$ is also manifest in the behavior of the hitting probability. As a function of the scaled target location $m$, the optimal rate $R^*(m)$ is also interesting (see \cite{supp}). Another observable that also confirms this optimal paradigm is the expected maximum of the RBB as a function of the rescaled resetting rate $R$ that we have computed exactly in \cite{supp} (as shown in the right panel in Fig.~\ref{fig:probh}).

To conclude, we derived an effective Langevin equation that generates, in a rejection-free manner, resetting Brownian bridge trajectories
in arbitrary dimensions. By computing analytically (i) the mean-square displacement, (ii) the hitting probability of a target and (iii) the expected maximum
of a one-dimensional resetting Brownian bridge, we have demonstrated that the paradigm of an optimal resetting rate exists, even in the presence of a bridge
constraint. Even though the value of the optimal resetting rate $r^*$ depends on the observables, they all exhibit a non-monotonic dependence on $r$ with
a single maximum. The presence of such an optimal resetting rate for bridges came out rather as a surprise and we have elucidated that the physical mechanism behind it is the result of a subtle
trade-off between resetting and the bridge constraint. This is very different from the physical mechanism that is responsible for an optimal $r^*$ for the standard resetting Brownian motion. We expect that this optimal paradigm exists for higher dimensional resetting Brownian bridges and possibly for other stochastic bridges -- it would be interesting to explore them in future studies.

\vspace*{0.3cm}

\acknowledgments{We thank H. Orland for useful discussions. This work was partially supported by the Luxembourg National Research Fund (FNR) (App. ID 14548297).}

%-----------------------------------------------------


\begin{thebibliography}{26}



\bibitem{Bartumeus09}
F. Bartumeus, J. Catalan, J. Phys. A: Math. Theor. \textbf{42}, 434002 (2009).

\bibitem{Viswanathan11}
G. M. Viswanathan, M. G. E. da Luz, E. P. Raposo, H. E. Stanley, \textit{The Physics of Foraging: An Introduction to Random Searches and Biological Encounters} (Cambridge: Cambridge
University Press, 2011).

\bibitem{Berg81}
O. G. Berg, R. B. Winter, P. H. von Hippel, Biochem. \textbf{20}, 6929 (1981).

\bibitem{Coppey04}
M. Coppey, O. B\'enichou, R. Voituriez, M. Moreau, Biophys. J. \textbf{87}, 1640 (2004).

\bibitem{Ghosh18}
S. Ghosh, B. Mishra, A. B. Kolomeisky, D. Chowdhury, J. Stat. Mech. 123209 (2018).

\bibitem{Chowdhury19}
D. Chowdhury, Biophys. J. \textbf{116}, 2057 (2019).

\bibitem{Adam68}
G. Adam, M. Delbr\"uck, \textit{Reduction of dimensionality in biological diffusion processes} in \textit{Structural Chemistry and Molecular Biology} (London: WH Freeman and Company, 1968).

\bibitem{Bell91}
W. J. Bell, \textit{ Searching Behaviour: The Behavioural Ecology of Finding Resources} (London:
Chapman and Hall, 1990).

\bibitem{Wolfe04}
J. M. Wolfe, T. S. Horowitz, Nat. Rev. Neurosci. \textbf{5}, 495 (2004).

 \bibitem{Montanari02}
A. Montanari, R. Zecchina, Phys. Rev. Lett. \textbf{88}, 178701 (2002).
 
   \bibitem{Oshanin07}
G. Oshanin, H. S. Wio, K. Lindenberg, S. F. Burlatsky, J. Phys. Condens. Matter \textbf{19}, 065142 (2007).
 
 \bibitem{Gelenbe10}
E. Gelenbe, Phys. Rev. E \textbf{82}, 061112 (2010).
 
 \bibitem{Snider12}
J. Snider, Phys. Rev. E \textbf{83}, 011105 (2011).
 
 \bibitem{Abdelrahman13}
O. H. Abdelrahman, E. Gelenbe, Phys. Rev. E \textbf{87}, 032125 (2013).
 

 
 \bibitem{Chupeau17}
M. Chupeau, O. B\'enichou, S. Redner, Phys. Rev. E \textbf{95}, 012157 (2017).
 

\bibitem{Lomholt08}
M. A. Lomholt, T. Koren, R. Metzler, J. Klafter, Proc. Natl. Acad. Sci. USA. \textbf{105}, 11055 (2008).

\bibitem{Benichou11}
O. B\'enichou, C. Loverdo, M. Moreau, R. Voituriez, Rev. Mod. Phys. \textbf{83}, 81 (2011).


 \bibitem{Evans20}
M. R. Evans, S. N. Majumdar, G. Schehr, J. Phys. A: Math. Theor. \textbf{53}, 193001 (2020).

\bibitem{Villen91}
M. Vill\'en-Altramirano, J. Vill\'en-Altramirano \textit{RESTART: a method for accelerating rare event
simulations} in \textit{Queueing Performance and Control in ATM} (Proc. 13th Int. Telegraphic Congress)
ed J. W. Cohen and C. D. Pack (Amsterdam: North-Holland, 1991).

\bibitem{Luby93}
M. Luby, A. Sinclair, D. Zuckerman, Inf. Process. Lett. \textbf{47}, 173 (1993).

\bibitem{Tong08}
H. Tong, C. Faloutsos, J-Y. Pan, Knowl. Inf. Syst. \textbf{14}, 327 (2008).

\bibitem{Avrachenkov13}
K. Avrachenkov, A. Piunovskiy, Y. Zhang, J. Appl. Probab. \textbf{50}, 960 (2013).

\bibitem{Lorenz18}
J. H. Lorenz \textit{Runtime distributions and criteria for restarts} in \textit{SOFSEM2018: Theory and Practice
of Computer Science} (Lecture Notes in Computer Science vol 10706) ed A. Tjoa et al (Cham: Edizioni della Normale, 2018) pp 493–507.

\bibitem{Reuveni14}
S. Reuveni, M. Urbakh, J. Klafter, Proc. Natl. Acad. Sci. USA. \textbf{111}, 4391 (2014).

\bibitem{Reuveni15}
T. Rotbart, S. Reuveni, M. Urbakh, Phys. Rev. E {\bf 92}, 060101 (2015).
%Michaelis-Mentenreactionschemeasaunifiedapproachtowards the optimal restart problem

\bibitem{Boyer14}
D. Boyer, C. Solis-Salas, Phys. Rev. Lett. {\bf 112}, 240601 (2014).
%Random walks with preferential relocations to places visited in the past and their application to biology

\bibitem{Majumdar15b}
S. N. Majumdar, S. Sabhapandit, G. Schehr, Phys. Rev. E \textbf{92}, 052126 (2015).

\bibitem{Boyer18}
G. Mercado-Vasquez, D. Boyer, J. Phys. A: Math. Theor. {\bf 51}, 405601 (2018).
%Lotka-Volterra systems with stochastic resetting




\bibitem{Maso19c}
A. Mas\'o-Puigdellosas, D. Campos, V. M\'endez, Front. Phys. \textbf{7}, 112 (2019).



\bibitem{Pal20}
A. Pal, L. Kusmierz, S. Reuveni, Phys. Rev. Research {\bf 2}, 043174 (2020). 
%2019 Home-range search provides advantage under high
%uncertainty 




\bibitem{Levikson77}
B. Levikson, J. Appl. Probab. \textbf{14}, 492 (1977).

\bibitem{Pakes78}
A. G. Pakes, J. Appl. Probab. \textbf{15}, 65 (1978).

\bibitem{Brockwell82}
P. J. Brockwell, J. Gani, S. I. Resnick, Adv. Appl. Probab. \textbf{14}, 709 (1982).

\bibitem{Brockwell85}
P. J. Brockwell, Adv. Appl. Prob. \textbf{17}, 42 (1985).

\bibitem{Kyriakidis94}
E. G. Kyriakidis, Stat. Probab. Lett. \textbf{20}, 239 (1994).

\bibitem{Pakes97}
A. G. Pakes, Commun. Stat. Stoch. Model. \textbf{13}, 255 (1997).

\bibitem{Manrubia99}
S. C. Manrubia, D. H. Zanette, Phys. Rev. E \textbf{59}, 4945 (1999).

\bibitem{Economou03}
A.  Economou, D. Fakinos, Eur. J. Oper. Res. \textbf{149}, 625 (2003).

\bibitem{Visco10}
P. Visco, R. J. Allen, S. N. Majumdar, M. R. Evans, Biophys. J. \textbf{98}, 1099 (2010).

\bibitem{Dharmaraja15}
S. Dharmaraja, A. Di Crescenzo, V. Giorno, A. G. Nobile, J. Stat. Phys. \textbf{161}, 326 (2015).

 \bibitem{Evans11}
M. R. Evans, S. N. Majumdar, Phys. Rev. Lett. \textbf{106}, 160601 (2011).
 
  \bibitem{RednerGuide}
S. Redner \textit{A guide to first-passage processes} (New York: Cambridge University Press, 2001).
  
\bibitem{Bray13}
A. J. Bray, S. N. Majumdar, G. Schehr, Adv. Phys. \textbf{62}, 225 (2013).

%%%%%%%%%%%%
  
  \bibitem{Evans11b}
M. R. Evans, S. N. Majumdar, J. Phys. A: Math. Theor. \textbf{44}, 435001 (2011).
 
 
 \bibitem{Evans14}
M. R. Evans, S. N. Majumdar, J. Phys. A: Math. Theor. \textbf{47}, 285001 (2014).
  
  \bibitem{Kusmierz14}
L. Ku\'smierz, S. N. Majumdar, S. Sabhapandit, G. Schehr, Phys. Rev. Lett. \textbf{113}, 220602 (2014).
  
  \bibitem{Christou15}
C. Christou, A. Schadschneider, J. Phys. A: Math. Theor. \textbf{48}, 285003 (2015). 
  
  \bibitem{Campos15}
D. Campos, V. M\'endez, Phys. Rev. E \textbf{92}, 062115 (2015).
  
  \bibitem{Montero16}
M. Montero, J. Villarroel, Phys. Rev. E \textbf{94}, 032132 (2016).
  
   
   \bibitem{Nagar16}
A. Nagar, S. Gupta, Phys. Rev. E \textbf{93}, 060102 (2016).
  
 \bibitem{Pal16}
A. Pal, A. Kundu, M. R. Evans, J. Phys. A: Math. Theor. \textbf{49}, 225001 (2016). 
  
 \bibitem{Eule16}
S. Eule, J. J. Metzger, New J. Phys. \textbf{18}, 033006 (2016).
 
  
 \bibitem{Reuveni16} 
S. Reuveni, Phys. Rev. Lett. {\bf 116}, 170601 (2016). 
%Optimalstochasticrestartrendersfluctuationsinfirstpassagetimesuniversal

\bibitem{Pal17}
A. Pal, S. Reuveni, Phys. Rev. Lett. {\bf 118}, 030603 (2017).
%First passage under restart 

\bibitem{Sokolov18}
A. Chechkin, I. M. Sokolov, Phys. Rev. Lett. {\bf 121}, 050601 (2018).  
 %Randomsearchwithresetting:aunifiedrenewalapproach 
  
  \bibitem{Evans18}
M. R. Evans, S. N. Majumdar, J. Phys. A: Math. Theor. \textbf{51}, 475003 (2018).

\bibitem{Prasad19a}
A. Pal, V. V. Prasad, Phys. Rev. Res. {\bf 1}, 032001 (2019).
%Landau-like expansion for phase transitions in stochastic resetting

\bibitem{Prasad19b}
A. Pal, V. V. Prasad, Phys. Rev. E {\bf 99}, 032123 (2019).
%First passage under stochastic resetting in an interval

\bibitem{Bressloff20a}
P. C. Bressloff, J. Phys. A: Math. Theor. {\bf 53}, 425001 (2020).
%Diffusive search for a stochastically-gated target with resetting.

\bibitem{Bressloff20b}
P. C. Bressloff, J. Phys. A: Math. Theor. {\bf 53}, 275003 (2020).
%Switching diffusions and stochastic resetting.
 
 \bibitem{DeBruyne20a} 
B. De Bruyne, J. Randon-Furling, S. Redner, Phys. Rev. Lett. \textbf{125}, 050602 (2020).

 \bibitem{DeBruyne20b} 
B. De Bruyne, J. Randon-Furling, S. Redner, J. Stat. Mech. 013203 (2020).

\bibitem{DeBruyneMori21} 
B. De Bruyne, F. Mori, arXiv:2112.11416.

\bibitem{Tal20}
O. Tal-Friedman, A. Pal, A. Sekhon, S. Reuveni, Y. Roichman, J. Phys. Chem. Lett. {\bf 11}, 7350 (2020).

\bibitem{Besga20}
B. Besga, A. Bovon, A. Petrosyan, S. N. Majumdar, S. Ciliberto, Phys. Rev. Res. {\bf 2}, 032029 (2020).

\bibitem{Faisant21}
F Faisant, B. Besga, A. Petrosyan, S. Ciliberto, S. N. Majumdar, J. Stat. Mech. 113203 (2021). 


\bibitem{Murphy92}
D. D. Murphy, B. R. Noon, \textit{Ecol. Appl.} \textbf{2}, 3 (1992).

\bibitem{Giuggioli05}
L. Giuggioli, G. Abramson, V. M. Kenkre, G. Suz\'an, E. Marc\'e, T. L. Yates, Bull. Math. Biol. \textbf{67}, 1135 (2005).

\bibitem{Shlesinger06}
M. F. Shlesinger, Nature \textbf{443}, 281 (2006).

\bibitem{Randon09}
J. Randon-Furling, S. N. Majumdar, A. Comtet, Phys. Rev. Lett. \textbf{103}, 140602 (2009).
 
 \bibitem{Boyle09}
S. A. Boyle, W. C. Lourenco, L. R. Da Silva, A. T. Smith, Folia Primatol. \textbf{80}, 33 (2009).

\bibitem{MajumdarCom10} 
S. N. Majumdar, A. Comtet, J. Randon-Furling, J. Stat. Phys. \textbf{138}, 955 (2010).
 
 \bibitem{Doob}
J. L. Doob, \textit{B. Soc. Math. Fr.} {\bf 85}, 431 (1957). 


\bibitem{Pitman}
P. Fitzsimmons, J. Pitman, M. Yor, {\it Seminar on Stochastic Processes} (Berlin: Springer, 1993).

\bibitem{BCDG2002}
P. G. Bolhuis, D. Chandler, C. Dellago, P. L. Geissler, Annu. Rev. Phys. Chem. {\bf 53}, 291 (2002).

\bibitem{GKP2006}
C. Giardin{\`a}, J. Kurchan, L. Peliti, Phys. Rev. Lett. {\bf 96}, 120603 (2006).

\bibitem{GKLT2011}
C. Giardin{\`a}, J. Kurchan, V. Lecomte, J. Tailleur, J. Stat. Phys. {\bf 145}, 787 (2011). 


\bibitem{Orland}
H. Orland, J. Chem. Phys. {\bf 134}, 174114 (2011).

\bibitem{CT2013}
R. Chetrite, H. Touchette, Ann. Henri Poincar\'e \textbf{16}, 2005 (2015). 

\bibitem{MajumdarEff15}
S. N. Majumdar, H. Orland, J. Stat. Mech. \textbf{6}, 06039 (2015).

\bibitem{Mazzolo17a}
A. Mazzolo, J. Stat. Mech., 023203 (2017).

\bibitem{Mazzolo17b}
A. Mazzolo, J. Math. Phys. \textbf{58}, 0953302 (2017).

\bibitem{KGGW2018}
K. Klymko, P. L. Geissler, J. P. Garrahan, S. Whitelam, Phys. Rev. E {\bf 97}, 032123 (2018). 

\bibitem{Gar2018}
J. P. Garrahan, \textit{Physica A} {\bf 504}, 130 (2018).

\bibitem{Brunet20}
E. Brunet, A. D. Le, A. H. Mueller, S. Munier, EPL {\bf 131}, 40002 (2020).
%How to generate the tip of branching random walks evolved to large times.

\bibitem{Rose21}
D. C. Rose, J. F. Mair, J. P. Garrahan, N. J. Phys. \textbf{23}, 013013 (2021).

\bibitem{Rose21area}
A. Das, D. C. Rose, J. P. Garrahan, D. T. Limmer, arXiv:2105.04321.

\bibitem{CLV21}
A. Chabane, A. Lazarescu, G. Verley, arXiv:2109.06830.


\bibitem{Baldassarri21}
A. Baldassarri, J. Stat. Mech. 083211 (2021).

\bibitem{Grela2021}
J. Grela, S. N. Majumdar, G. Schehr, J. Stat. Phys. \textbf{183}, 1 (2021).


\bibitem{DebruyneAR21}
B. De Bruyne, S. N. Majumdar, H. Orland, G. Schehr, J. Stat. Mech., 123204 (2021).


\bibitem{Monthus21}
C. Monthus, arXiv:2111.05696.

\bibitem{DebruyneRW21}
B. De Bruyne, S. N. Majumdar, G. Schehr, Phys. Rev. E. \textbf{104}, 024117 (2021).

\bibitem{DebruyneRTP21}
B. De Bruyne, S. N. Majumdar, G. Schehr, J. Phys. A: Math. Theor. \textbf{54}, 385004 (2021).

\bibitem{RG2017}
E. Roldan, S. Gupta, Phys. Rev. E {\bf 96}, 022130 (2017).
%Path-integral formalism for stochastic resetting: Exactly solved examples and shortcuts to confinement

\bibitem{Shk17}
V. P. Shkilev, Phys. Rev. E {\bf 96}, 012126 (2017).
%Continuous-time random walk under time-dependent resetting


\bibitem{Pinsky20}
R. G. Pinsky, Stoch. Proc. Appl. {\bf 130}, 2954 (2020).
%Diffusive search with spatially dependent resetting. 

\bibitem{Ray20}
S. Ray, J. Chem. Phys. {\bf 153}, 234904 (2020).
%Space-dependent diffusion with stochastic resetting: A first-passage study


\bibitem{BS20}
A. S. Bodrova, I. M. Sokolov, Phys. Rev. E {\bf 102}, 032129 (2020).
%Brownian motion under non-instantaneous resetting in higher dimensions

\bibitem{supp}
See Supplemental Material at [URL will be inserted by publisher] for detailed computations.

\end{thebibliography}
\end{document}

%\preprint{}
\title{Optimal Resetting Brownian Bridges: Supplemental Material}% Force line breaks with \\
\author{Benjamin De Bruyne}
\affiliation{LPTMS, CNRS, Univ.  Paris-Sud,  Universit\'e Paris-Saclay,  91405 Orsay,  France}
\author{Satya N. Majumdar}
\affiliation{LPTMS, CNRS, Univ.  Paris-Sud,  Universit\'e Paris-Saclay,  91405 Orsay,  France}
\author{Gr\'egory Schehr}
\affiliation{Sorbonne Universit\'e, Laboratoire de Physique Th\'eorique et Hautes Energies, CNRS UMR 7589, 4 Place Jussieu, 75252 Paris Cedex 05, France}

%\date{\today}

\begin{abstract}
We give the principal details of the calculations described in the main text of the Letter. 
\end{abstract}

\maketitle

%\tableofcontents
\newpage
\section{Derivation of the effective Langevin equation in $d$ dimensions}
In this section, we give the details of the derivation of the effective Langevin equation for the resetting Brownian bridge (RBB) to $d$ dimensions. We perform the computation in the general setup where the initial and final positions are different:
\begin{align}
  \mathbf{x}_B(0)=\mathbf{0}\,,\quad \mathbf{x}_B(t_f)=\mathbf{x}_f\,.\label{eq:const}
\end{align}
The effective equation of motion for a RBB can be straightforwardly generalized in $d$ dimensions. It writes:
\begin{align}
   \mathbf{x}_B(t+dt) = \left\{\begin{array}{lll}
    \mathbf{x}_B(t) + \sqrt{2D}\,\bm{\eta}(t) dt +  \bm{\tilde\mu}(\mathbf{x}_B,t) d t &\text{with prob.} &1-\tilde r(\mathbf{x}_B,t)\,dt\,,\\
    \mathbf{x}_B(0)=0&\text{with prob.} &\tilde r(\mathbf{x}_B,t)\,dt\,,
  \end{array}\right.\label{eq:eomeffd}
\end{align}
where the effective drift $\bm{\tilde\mu}$ and resetting rate $\tilde r$ are given by 
\begin{subequations}
\begin{align}
 \bm{\tilde\mu}(\mathbf{x},t) &= 2D\nabla_\mathbf{x} \ln(Q_r(\mathbf{x},t)) \,,\label{eq:effdd}\\
   \tilde r(\mathbf{x},t) &= r \,\frac{Q_r(\mathbf{0},t)}{Q_r(\mathbf{x},t)}\,,\label{eq:effrd}
\end{align}
\end{subequations}
where $Q_r(\mathbf{x},t)\equiv P_r(\mathbf{0},t_f\,|\,\mathbf{x},t)$ is the backward propagator of resetting Brownian motion (without constraints). To derive the evolution equation for the backward propagator $Q_r(\mathbf{x},t)$, we consider evolving the process from $t$ to $t+ dt$. In this small time interval $dt$, we see from the equation of motion in Eq.~(1) in the main text that the position either (i) moved from ${\bm x}$ to ${\bm x}+\sqrt{2D}{\bm\eta}(t)\,dt$ with probability $1-rdt$ or (ii) reset from ${\bm x}$ to ${\bm 0}$ with probability $rdt$. For the subsequent evolution from $t+dt$ to $t_f$, the ``new initial position'' of the process is therefore ${\bm x}+\sqrt{2D}{\bm\eta}(t)dt$ with probability $1-rdt$ or ${\bm 0}$ with probability $rdt$. Averaging over all possible values of ${\bm\eta}(t)$, we obtain
\begin{align}
  Q_r(\mathbf{x},t) = (1-rdt)  \langle Q_r({\bm x}+\sqrt{2D}\,{\bm\eta}(t)dt,t+dt) \rangle_{\bm \eta} + rdt \,Q_r(\mathbf{0},t+dt)\,. \label{eq:Qdt}
\end{align} 
Taking the limit $dt\to 0$, we obtain the backward Fokker-Plank equation
\begin{align}
 -\partial_t Q_r(\mathbf{x},t) = \Delta Q_r(\mathbf{x},t) - r Q_r(\mathbf{x},t) + r \,Q_r(\mathbf{0},t)\,,\label{eq:Qrd}
\end{align}
with the final condition $Q_r(\mathbf{x},t_f)=\delta(\mathbf{x}-\mathbf{x}_f)$. One can similarly derive the equation for the forward propagator $P_r(\mathbf{x},t)\equiv P_r(\mathbf{x},t\,|\,\mathbf{0},0)$ as follows. We again evolve the joint
process from $t$ to $t + dt$. In this interval, we see from the equation of motion that the particle either (i)
 traveled from ${\bm x}-\sqrt{2D}{\bm\eta}(t)\,dt$ to ${\bm x}$ with probability $1-rdt$ or (ii) reset from any $\mathbf{x}'$ to $\mathbf{0}$ with probability $rdt$. Averaging over all possible values of ${\bm\eta}(t)$, we obtain
\begin{align}
  P_r(\mathbf{x},t) = (1-rdt)  \langle P_r({\bm x}-\sqrt{2D}\,{\bm\eta}(t)dt,t-dt) \rangle_{\bm \eta} + rdt \,\delta(\mathbf{x})\int_{-\infty}^\infty d\mathbf{x'}\,P_r(\mathbf{x'},t-dt)\,. \label{eq:Pdt}
\end{align} 
Taking the limit $dt\to 0$ and using the normalization $\int_{-\infty}^\infty d\mathbf{x'}\,P_r(\mathbf{x'},t)=1$, we obtain the forward Fokker-Plank equation
\begin{align}
 \partial_t P_r(\mathbf{x},t) = \Delta P_r(\mathbf{x},t) - r P_r(\mathbf{x},t) + r\, \delta(\mathbf{x})\,,\label{eq:Prd}
\end{align}
with the initial condition $P_r(\mathbf{x},0)=\delta(\mathbf{x})$.
The expressions in Eq.~(\ref{eq:Qrd}) and Eq.~(\ref{eq:Prd}) in $d=1$ recover the ones given in Eqs.~(6) in the main text.
 We now calculate the backward propagator $Q_r(\mathbf{x},t)$ explicitly. To do so, one can solve the backward Fokker-Plank equation in Eq.~(\ref{eq:Qrd}), or alternatively rely on the renewal identity for resetting \cite{Evans20}
\begin{align}
  P_r(\mathbf{x},t\,|\,\mathbf{x}_0,0) = e^{-rt}P_0(\mathbf{x},t\,|\,\mathbf{x}_0,0) + r \int_0^t d\tau e^{-r\tau}P_0(\mathbf{x},\tau\,|\,\mathbf{0},0)\,,\label{eq:rend}
\end{align}
where $P_0(\mathbf{x},t\,|\,\mathbf{x}_0,0)=e^{-\frac{(\mathbf{x}-\mathbf{x}_0)^2}{4Dt}}/(4\pi Dt)^{d/2}$ is the free Brownian propagator \cite{RednerGuide,Bray13}, which gives
\begin{align}
  Q_r(\mathbf{x},t) &\equiv P_r(\mathbf{x}_f,t_f\,|\,\mathbf{x},t) =e^{-r(t_f-t)}\frac{e^{-\frac{(\mathbf{x}_f-\mathbf{x})^2}{4 D (t_f-t)}}}{(4\pi D (t_f-t))^{d/2}}+ r \int_0^{t_f-t} d\tau e^{-r\tau} \frac{e^{-\frac{\mathbf{x}_f^2}{4 D \tau}}}{(4\pi D \tau)^{d/2}}\,.\label{eq:Qd}
\end{align}
Substituting this expression (\ref{eq:Qd}) in Eqs. (\ref{eq:effdd}) and (\ref{eq:effrd}), we get the effective drift and resetting rates 
\begin{subequations}
\begin{align}
\tilde{\bm{\mu}}(\mathbf{x},t) &= -\sqrt{4 \,r D} \,\bm{V}_d\left(\mathbf{y}=\frac{\mathbf{x}}{\sqrt{4D(t_f-t)}},z=r(t_f-t),\mathbf{y}_f=\frac{\mathbf{x}_f}{\sqrt{4D(t_f-t)}} \right)\,,\label{eq:effmusd}\\
  \tilde r(\mathbf{x},t) &= r\, \mathcal{W}_d\left(\mathbf{y}=\frac{\mathbf{x}}{\sqrt{4D(t_f-t)}},z=r(t_f-t),\mathbf{y}_f=\frac{\mathbf{x}_f}{\sqrt{4D(t_f-t)}}\right)\,,\label{eq:effrsd}
\end{align}
\label{eq:murd}
\end{subequations}
where 
\begin{subequations}
\begin{align}
\bm{V}_d(\mathbf{y},z,\mathbf{y}_f) &= \frac{(\mathbf{y}_f-\mathbf{y})\, e^{-(\mathbf{y}_f-\mathbf{y})^2 -z}}{\sqrt{z}\left[e^{-z-(\mathbf{y}_f-\mathbf{y})^2}+z \int_0^1 du \frac{1}{u^{d/2}}\, e^{-zu-\frac{\mathbf{y}_f^2}{u}} \right]}\,, \label{eq:Vd}\\
  \mathcal{W}_d(\mathbf{y},z,\mathbf{y}_f) &= \frac{e^{-z-\mathbf{y}_f^2} + z \int_0^1 du \frac{1}{u^{d/2}}\, e^{-zu-\frac{\mathbf{y}_f^2}{u}} }{e^{-z-(\mathbf{y}_f-\mathbf{y})^2}+z \int_0^1 du \frac{1}{u^{d/2}}\, e^{-zu-\frac{\mathbf{y}_f^2}{u}} }\,.\label{eq:Wd}
\end{align}
\label{eq:VdWd}
\end{subequations}
The effective Langevin equation in Eq.~(\ref{eq:eomeffd}) along with the effective drift and resetting rate above can then be used to generate RBB in arbitrary dimensions. When $d=1$ and $x_f=0$, we recover the effective drift and resetting rate displayed in Eqs. (4a) and (4b) in  the main text. \\

\vspace*{0.5cm}
\noindent{\bf The case $d\geq 2$}. From the formulae in Eq. (\ref{eq:VdWd}), one sees that the integrals appearing in the denominator are convergent for $d \geq 2$ only when the rescaled final position ${\bf y}_f \neq {\bf 0}$. If ${\bf y}_f$ is set to be ${\bf y}_f= {\bf 0}$, this divergence in the denominator leads to an effective drift that is exactly zero and a resetting rate which is equal to $r$ for $t<t_f$ and is infinite for $t=t_f$ (note that for $t=t_f$, $z=0$ and $|{\bf y}| \to \infty$). Therefore, in $d \geq 2$, a RBB with $\mathbf{x}_f=\mathbf{0}$ can be simply generated by a free RBM which is forced to reset at the last instant $t=t_f$. In contrast, 
when $\mathbf{x}_f\neq\mathbf{0}$ in $d\geq 2$, $\tilde{\bm{\mu}}(\mathbf{x},t) $ and $\tilde r(\mathbf{x},t)$ are both finite and one needs to use their full expressions given in Eqs.~(\ref{eq:VdWd}) to generate a RBB. The peculiarity observed when $\mathbf{x}_f=\mathbf{0}$ in $d\geq 2$ can be further investigated by considering the limit $|\mathbf{x}_f|\to 0$ and $t_t-t\to 0$ while keeping $\mathbf{y}_f=\frac{\mathbf{x}_f}{\sqrt{4D(t_f-t)}}$ fixed in Eqs.~(\ref{eq:murd}). We observe that when $\sqrt{t-t_f}\gg |\mathbf{x}_f| $, the effective drift and resetting rate tend towards their singular values, while when $\sqrt{t-t_f}\ll |\mathbf{x}_f| $, i.e.,~close to the end of the bridge, the effective drift and resetting rate take finite values given by their full expressions in Eqs.~(\ref{eq:VdWd}).

\section{Exact computation of the search efficiency}

In this section, we compute three different observables for the RBB in $d=1$ with $x_B(0)=x_B(t_f)=0$: (i) the mean-square displacement, (ii) the hitting probability of a target at a fixed location $M$ and (iii) the expected maximum of the RBB. All the three observables can be used as indicators of the efficiency of the search process via a RBB and all of them exhibit an optimal resetting rate.

\subsection{Mean square displacement}
In this section, we compute the mean square displacement of a RBB at an intermediate time $t\leq t_f$. The mean square displacement $\langle x_B^2\rangle (t|t_f)$  is given by 
\begin{equation}
\langle x_B^2\rangle (t|t_f)= \int_{-\infty}^{\infty} dx\, x^2\, P_B(x,t|t_f),
\label{msd_rbb.1}
\end{equation}
where $P_B(x,t|t_f)$ is the bridge probability distribution which, as it was shown in the main text, is given by 
\begin{align}
   P_B(x,t\,|\,t_f) =\frac{P_r(x,t\,|\,0,0)P_r(0,t_f\,|\,x,t)}{P_r(0,t_f\,|\,0,0)},\label{eq:Pb}
\end{align}
where $P_r(x,t\,|\,x_0,t_0)$ is the propagator of the RBM given by the renewal identity in Eq.~(\ref{eq:rend}) with $d=1$. This renewal identity gives the following expressions for the two terms in the numerator in Eq.~(\ref{eq:Pb}):
\begin{align}
  P_r(x,t\,|\,0,0) &= e^{-rt}\frac{e^{-\frac{x^2}{4 D t}}}{\sqrt{4\pi } \sqrt{D t}}+ \frac{1}{4} \sqrt{\frac{r}{D}} e^{-| x| \sqrt{\frac{r}{D}}} \text{erfc}\left(\frac{| x| -t \sqrt{4D r}}{
   \sqrt{4D t}}\right)-\frac{1}{4} \sqrt{\frac{r}{D}} e^{| x|  \sqrt{\frac{r}{D}}} \text{erfc}\left(\frac{| x| + t \sqrt{4D r}}{ \sqrt{4D t}}\right)  \,,\label{eq:P}\\
  P_r(0,t_f\,|\,x,t) &=\sqrt{\frac{r}{4D}} \text{erf}\left(\sqrt{r \left(t_f-t\right)}\right)+e^{-r \left(t_f-t\right)}\frac{e^{-\frac{x^2}{4 D
   \left(t_f-t\right)}}}{\sqrt{4\pi } \sqrt{D \left(t_f-t\right)}}\,,\label{eq:Q}
\end{align}
where ${\rm erfc}(z) = 1 - {\rm erf}(z) = (2/\sqrt{\pi})\int_z^{\infty} e^{-u^2}\, du$ is the complementary error function. The denominator in Eq.~(\ref{eq:Pb}) then reads
\begin{align}
  P_r(0,t_f\,|\,0,0)=\sqrt{\frac{r}{4D}} \text{erf}\left(\sqrt{r t_f}\right)+\frac{e^{-r t_f}}{\sqrt{4\pi D t_f}}\,. \label{eq:den}
\end{align} 
Inserting Eqs.~(\ref{eq:P}), (\ref{eq:Q}) and (\ref{eq:den}) into Eq.~(\ref{eq:Pb}), we find that the full expression of the RBB 
propagator 
\begin{align}
  P_B(x,t\,|\,t_f) &=  \frac{\sqrt{\pi  t_f}}{\sqrt{\pi r t_f}\,\text{erf}(\sqrt{r t_f})+e^{-r t_f}}\left[ \sqrt{r} \text{erf}\left(\sqrt{r \left(t_f-t\right)}\right)+e^{-r \left(t_f-t\right)}\frac{e^{-\frac{x^2}{4 D
   \left(t_f-t\right)}}}{\sqrt{\pi\left(t_f-t\right)}}\right]\nonumber\\
   &\times\left[e^{-rt}\frac{e^{-\frac{x^2}{4 D t}}}{\sqrt{\pi t}}+ \frac{1}{2} \sqrt{r} e^{-| x| \sqrt{\frac{r}{D}}} \text{erfc}\left(\frac{| x| -t \sqrt{4D r}}{
   \sqrt{4D t}}\right)-\frac{1}{2} \sqrt{r} e^{| x|  \sqrt{\frac{r}{D}}} \text{erfc}\left(\frac{| x| + t \sqrt{4D r}}{ \sqrt{4D t}}\right) \right]\,.\label{eq:Pbfull}
\end{align}
A plot of $P_B(x,t\,|\,t_f)$ is given in Fig.~1 (right panel) in the main text. When $r=0$, we recover the propagator of the Brownian bridge \cite{Rogers00} 
\begin{align}
  P_B(x,t\,|\,t_f)=\frac{\sqrt{t_f}}{\sqrt{4 \pi Dt(t_f-t)}}e^{-\frac{t_f x^2}{4 Dt(t_f-t)}}\,.\label{eq:PbBB}
\end{align} 

The mean square displacement in Eq.~(\ref{msd_rbb.1}) can in principle be computed exactly by using Eq.~(\ref{eq:Pbfull}) in (\ref{msd_rbb.1}). However this integral is 
a bit cumbersome to perform. Hence, we use an alternative approach, relying on the renewal equation (\ref{eq:rend}) which gives
\begin{align}
  \langle x_B^2\rangle (t|t_f)= \frac{1}{P_r(0,t_f\,|\,0,0)}\int_{-\infty}^\infty dx &\,x^2 \left[e^{-rt}P_0(x,t\,|\,0,0) + r \int_0^t d\tau_1 e^{-r\tau_1}P_0(x,\tau_1\,|\,0,0)\right]\nonumber\\
 & \times \left[e^{-r(t_f-t)}P_0(0,t_f-t\,|\,x,0) + r \int_0^{t_f-t} d\tau_2 e^{-r\tau_2}P_0(0,\tau_2\,|\,0,0)\right]\,.
\end{align}
By further using the expression of the Gaussian propagator $P_0(x,t\,|\,x_0,0)=e^{-\frac{(x-x_0)^2}{4Dt}}/(4\pi Dt)^{1/2}$,  one finds 
\begin{equation}
\langle x_B^2\rangle (t|t_f)= 2D\, t_f \, f(a|R)\, , \quad {\rm where}\quad a=\frac{t}{t_f} \in [0,1] \quad {\rm and}\quad R= r\, t_f\,.
\label{msd_rbb.2}
\end{equation}
The scaling function $f(a|R)$ is given by
\begin{equation}
f(a|R)= \frac{N(a|R)}{\left[e^{-R}+ 
\sqrt{\pi R}\,{\rm erf}\left(\sqrt{R}\right)\right]}
\, ,
\label{far.0}
\end{equation}
where ${\rm erf}(z) = (2/\sqrt{\pi})\int_0^z e^{-u^2}\,du$ and $N(a|R)$ reads
\begin{align}
  N(a|R) &=a(1-a)e^{-R} + a R \int_0^{1-a}d\tau_2\, e^{-R(a+\tau_2)} \frac{1}{\sqrt{\tau_2}} + (1-a)R\int_0^{a}d\tau_1\, e^{-R(1-a+\tau_1)} \frac{\tau_1}{(1-a+\tau_1)^{3/2}}\nonumber\\
&  \quad + R^2 \int_0^{1-a}d\tau_2\int_0^{a}d\tau_1\, e^{-R(\tau_2+\tau_1)} \frac{\tau_1}{\sqrt{\tau_2}}\,.\label{eq:NaR2}
\end{align}
The function $N(a|R)$ can be further simplified by performing the integrals in Eq.~(\ref{eq:NaR2}) and gives
\begin{align}
  N(a|R) &= (1-a)[a+2(1-a)R]\,e^{-R}-2(1-a)^{3/2}R\,e^{-R(1-a)}+\sqrt{\frac{\pi}{R}}\left(1-e^{-aR}\right)\,\text{erf}\left(\sqrt{R(1-a)}\right)\nonumber \\
  &\quad +\sqrt{\pi R}(1-a)[1+2(1-a)R]\left[\text{erf}\left(\sqrt{R}\right)-\text{erf}\left(\sqrt{R(1-a)}\right)\right]\,.\label{eq:NaR}
\end{align}
The expression in Eq.~(\ref{eq:NaR}) is exact for all $R$. Upon taking the limit $R\to 0$ and $R\to \infty$, we recover the asymptotic expressions given in the main text below Eq.~(10).

\subsection{Hitting probabilty}

\begin{figure}[t]
\center
 \includegraphics[width=0.4\textwidth]{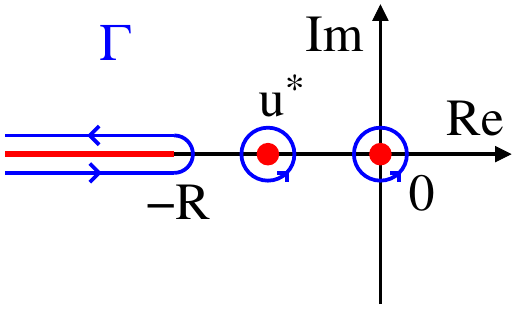}%
\caption{Integration contour $\Gamma$ to compute the integral in Eq.~(\ref{eq:hRy}). The integrand has two simple poles located at $u=0$ and at $u^*>-R$ along with a branch cut attached to $u=-R$. The integration contour $\Gamma$ is set to go around the poles and along the branch cut. }
\label{fig:contour}
\end{figure}

\begin{figure}[t]
\center
 \includegraphics[width=0.4\textwidth]{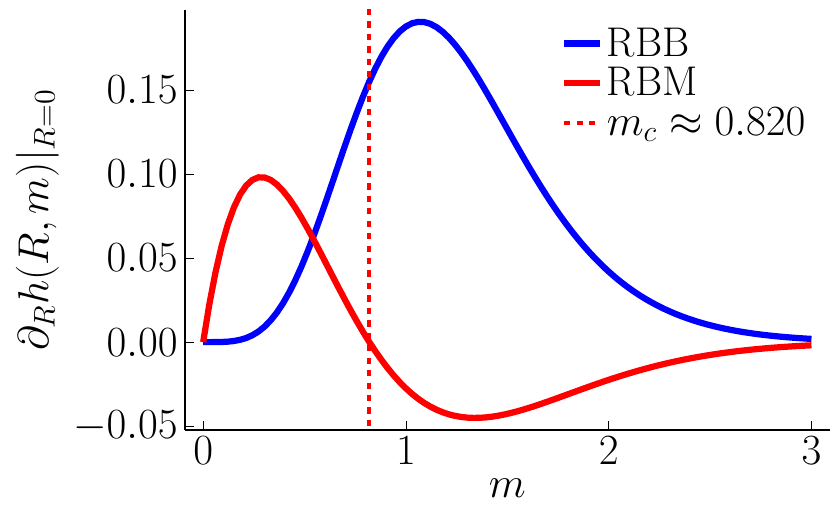}%
\caption{Slope of the scaling function $h(R,m)$ at $R=0$ given by the linear term in $R$ in Eq.~(\ref{eq:hr0a}) for the RBB (blue curve) and in Eq.~(\ref{eq:hnoR0}) for the RBM (red curve) as a function of $m$. As can be seen, while the slope is also positive for the RBB, the slope for the RBM becomes negative beyond a critical value $m=m_c\approx 0.8198$. This induces a freezing of the optimal resetting rate (see Fig.~\ref{fig:roptz}).}
\label{fig:dh}
\end{figure}

\begin{figure}[t]
\center
\includegraphics[width=0.4\textwidth]{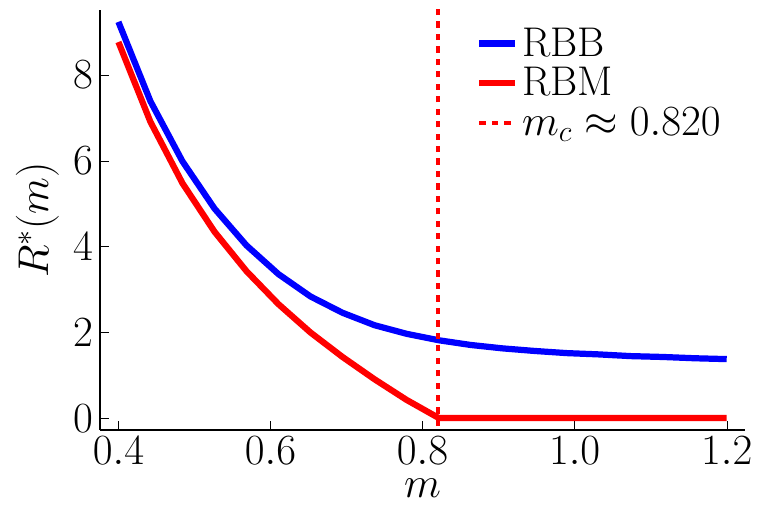}
\caption{Optimal resetting rate $R^*(m)$ to maximize the probability to reach $m=M/\sqrt{2Dt_f}$ for the RBB (blue curve) and the RBM (red curve). The optimal resetting rate has been obtained numerically by maximizing Eq.~(\ref{eq:h2}) for the RBB and Eq.~(\ref{eq:hRyn}) for the RBM. The optimal resetting rate for the RBM freezes to $0$ beyond a critical value $m=m_c\approx 0.820$.  }
\label{fig:roptz}
\end{figure}

In this section, we compute the hitting probability of a target located at $M>0$ by the RBB of fixed duration $t_f$ and with $x_B(0) = x_B(t_f) = 0$. As explained in the main text,
the hitting probability $p_{\rm hit}(t_f,M)$ can be expressed in terms of the first-passage probability density $F_B(t\,|\,M,t_f)$ via the relation 
\begin{align}
  p_{\text{hit}}(t_f,M) = \int_0^{t_f} dt F_B(t\,|\,M,t_f)\,.\label{eq:phit}
\end{align}
We first compute $F_B(t\,|\,M,t_f)$, i.e., the probability distribution of the time $t$ at which a RBB will reach the level $M$ for the first time given that it must return to the origin at a future time $t_f$. As it was shown in the main text, $F_B(t\,|\,M,t_f)$ is given by 
\begin{align}
  F_B(t\,|\,M,t_f) = \frac{F_r(t\,|\,M) P_r(0,t_f\,|\,M,t)}{P_r(0,t_f\,|\,0,0)}\,,\label{eq:FB}
\end{align}
where $F_r(t\,|\,M)$ is the first-passage time distribution of the RBM \cite{Evans11}, $P_r(x,t\,|\,x_0,t)$ is the propagator of the RBM given by the renewal identity in Eq.~(\ref{eq:rend}) with $d=1$.
The first-passage distribution of resetting Brownian motion is given by \cite{Evans11}
\begin{align}
 F_r(t\,|\,M) = \int_\Gamma \frac{ds}{2\pi i}\,e^{s\,t} \,\frac{r+s}{r+s\, e^{\sqrt{\frac{r+s}{D}}M} }\,,\label{eq:FrM}
\end{align}
where $\Gamma$ is the usual Bromwich contour. Upon inserting Eq.~(\ref{eq:P}), (\ref{eq:den}) and (\ref{eq:FrM}) in Eq.~(\ref{eq:FB}), we find that first-passage time distribution $F_B(M,t\,|\,t_f)$ is given by
\begin{align}
   F_B(t\,|\,M,t_f) = \frac{1}{t_f}\, g(a\,|\,m,R)\,,\quad {\rm where}\quad a=\frac{t}{t_f}\,, \quad m=\frac{M}{\sqrt{2Dt_f}}\quad  {\rm and}\quad R= r\, t_f\,.\label{eq:FB2}
\end{align}
The scaling function $g(a\,|\,m,R)$ is given explicitly by
\begin{align}
  g(a\,|\,m,R) = \frac{1}{\sqrt{(1-a)}}\,\frac{\sqrt{\pi R(1-a)}\,\text{erf}\left(\sqrt{\pi R(1-a)}\right)+e^{-R(1-a)-\frac{m^2}{2(1-a)}}}{\sqrt{\pi R}\,\text{erf}\left(\sqrt{R}\right)+e^{-R}}\int_\Gamma \frac{du}{2\pi i} e^{u\,a}\,\frac{R+u}{R+u\,e^{m\sqrt{2}\sqrt{R+u}}}\,.\label{eq:g}
\end{align}
For $R=0$, we recover the standard result for a Brownian bridge \cite{Billingsley68}
\begin{align}
   g(a\,|\,m,R=0) = \frac{m}{\sqrt{2\pi a^3(1-a)}}\,e^{-\frac{m^2}{2(1-a)a}}\,.\label{eq:g0}
\end{align}
We now compute the hitting probability by inserting Eq.~(\ref{eq:FB}) into (\ref{eq:phit}), which gives
\begin{align}
  p_{\text{hit}}(t_f,M) &= \frac{1}{P_r(0,t_f\,|\,0,0)} \int_0^{t_f}dt F_r(t\,|\,M) P_r(0,t_f\,|\,M,t) \,,\nonumber \\
   & =\frac{\sqrt{\pi  t_f}}{\sqrt{\pi r t_f}\,\text{erf}(\sqrt{r t_f})+e^{-r t_f}}\, I(t_f,M)\,,\label{eq:phi}
\end{align}
where we used the expression of $P_r(0,t_f\,|\,0,0)$ given in Eq.~(\ref{eq:den}) and denoted 
\begin{align}
   I(t_f,M) = \int_0^{t_f}dt F_r(t\,|\,M) P_r(0,t_f\,|\,M,t) \,.\label{eq:I}
\end{align}
Since $P_r(0,t_f\,|\,M,t)$ in Eq.~(\ref{eq:I}) only depends on the difference $t_f-t$ [see Eq.~(\ref{eq:Q})], the integral in Eq.~(\ref{eq:I}) has a convolution structure and it is convenient to consider its Laplace transform
\begin{align}
 \tilde I(s,M)= \int_0^\infty dt_f\, e^{-st_f} I(t_f,M) =\frac{\sqrt{s+r}}{2s\sqrt{D}}\left[\frac{r+s\,e^{-\sqrt{\frac{r+s}{D}}M}}{r+s\, e^{\sqrt{\frac{r+s}{D}}M} }\right]\,,\label{eq:IsM}
\end{align}
where we used the Laplace transform of $F_r(t\,|\,M)$ given in Eq.~(\ref{eq:FrM}) and the Laplace transform of $P_r(0,t_f\,|\,M,0)$ given by $\int_0^\infty dt_f\, e^{-st_f}P_r(0,t_f\,|\,M,0)=\frac{1}{2\sqrt{D(r+s)}} \left(\frac{r}{s}+e^{-\sqrt{\frac{r+s}{D}}M}\right)$. Formally inverting the Laplace transform in Eq.~(\ref{eq:IsM}), we find
\begin{align}
  I(t_f,M) = \int_\Gamma \frac{ds\,e^{s t_f}}{2\pi i}\,\frac{\sqrt{s+r}}{2s\sqrt{D}}\left[\frac{r+s\,e^{-\sqrt{\frac{r+s}{D}}M}}{r+s\, e^{\sqrt{\frac{r+s}{D}}M} }\right]\,,\label{eq:ItfM}
\end{align}
where $\Gamma$ is the usual Bromwich contour.
Upon inserting Eq.~(\ref{eq:ItfM}) into Eq.~(\ref{eq:phi}), we obtain
\begin{align}
   p_{\text{hit}}(t_f,M) &= h\left(R=rt_f,m=\frac{M}{\sqrt{2D t_f}}\right)\,,\label{eq:phits}
\end{align}
where the scaling function is
\begin{align}
  h(R,m) = \frac{\sqrt{\pi }}{\left(\sqrt{\pi R}\,\text{erf}\left(\sqrt{R}\right)+e^{-R}\right)}\,\int_\Gamma \frac{du\,e^{u}}{2\pi i}  \frac{\sqrt{u+R}}{u}\,\frac{R+u \,e^{-m\sqrt{2}\sqrt{u+R}}}{R+u \,e^{m\sqrt{2}\sqrt{u+R}}}\,.\label{eq:hRy}
\end{align}
When $R=0$, we recover the Brownian bridge result  (see e.g. \cite{Rogers00})
\begin{align}
  h(R=0,m) = e^{-2m^2} \,.\label{eq:hR0}
\end{align}
One can further simplify the expression (\ref{eq:hRy}) by examining the singular structure of the integrand in the complex $u$ plane. It has two simple poles located at $u=0$ and at $u^*>-R$ which satisfies
\begin{align}
 R+u^* \,e^{m\sqrt{2}\sqrt{u^*+R}} = 0 \,,\label{eq:us}
\end{align}
along with a branch cut attached to $u=-R$ (see Fig.~\ref{fig:contour}).
 By performing the contour integration along the branch cut and the two poles, we find
\begin{align}
  h(R,m) = 1 - \frac{\sqrt{\pi}}{\sqrt{\pi R}\,\text{erf}(\sqrt{R})+e^{-R}}\Bigg[&\frac{\sqrt{2}e^{u^*}(u^*+R)(R^2-u^{*2})}{R^2[\sqrt{2}\sqrt{u^*+R}+u^*m]}+ 2 \int_0^\infty \frac{du}{\pi}\frac{\sqrt{u}\,e^{-(R+u)}(R+u)\,\sin^2(\sqrt{2u}m)}{2R(R+u)[1-\cos(\sqrt{2u}m)]+u^2}\Bigg]\,.\label{eq:h2}
\end{align}
Despite the presence of oscillatory functions in the integrand in Eq.~(\ref{eq:h2}), $h(R,m)$ does not display any oscillations as a function of $m$. For $R\to \infty$, the solution of the non-linear equation in Eq.~(\ref{eq:us}) is $
  u^* \sim -R\,e^{-m\,\sqrt{2 R}}$
and we get 
\begin{align}
   h(R,m) \sim R \,e^{-\sqrt{2} \sqrt{R} m}\,,\quad R\to \infty\,.\label{eq:hinf}
\end{align}
For $R\to 0$, the solution of the non-linear equation in Eq.~(\ref{eq:us}) is $
  u^* \sim -R + 2m^2 R^2$ and we find
\begin{align}
 h(R,m)\sim  e^{-2m^2}+ R\left[\sqrt{2 \pi } m
   \left(\text{erf}\left(\frac{m}{\sqrt{2}}\right)-3\,
   \text{erf}\left(\frac{3 m}{\sqrt{2}}\right)+2\,
   \text{erf}\left(\sqrt{2} m\right)\right)-2\,
   e^{-\frac{9 m^2}{2}}+2 \, e^{-\frac{m^2}{2}}\right]\,,\quad R\to 0\,.\label{eq:hr0a}
\end{align}
 The term in brackets in Eq.~(\ref{eq:hr0a}) is always positive (see blue curve in Fig.~\ref{fig:dh}), which indicates that the function $h(R,m)$ as a function of $R$ has a non-monotonic behavior for all $m$ and is maximized at some value $R^*(m)$.
The optimal value $R^*(m)$ as a function of the rescaled target
distance $m$ decays monotonically with increasing $m$ (see
blue curve in Fig.~\ref{fig:roptz}). This behavior is quite different from the resetting Brownian motion without the bridge constraint for which the hitting probability $ p_{\text{hit, no bridge}}(t_f,M)$ is given by $p_{\text{hit, no bridge}}(t_f,M)=\int_0^t F_r(t\,|\,M)$, where $F_r(t\,|\,M)$ is the first-passage distribution given in Eq.~(\ref{eq:FrM}). Using the expression in Eq.~(\ref{eq:FrM}), we obtain the hitting probability 
\begin{align}
   p_{\text{hit, no bridge}}(t_f,M) &= h_{\text{no bridge}}\left(R=rt_f,m=\frac{M}{\sqrt{2D t_f}}\right)\,,\label{eq:phitsno}
\end{align}
where
\begin{align}
  h_{\text{no bridge}}(R,m) = \int_\Gamma \frac{du\,e^{u}}{2\pi i}  \frac{1}{u}\,\frac{R+u }{R+u \,e^{m\sqrt{2}\sqrt{u+R}}}\,.\label{eq:hRyn}
\end{align}
The function $h_{\text{no bridge}}(R,m)$ behaves for $R\to 0$ as
\begin{align}
 h_{\text{no bridge}}(R,m)  &\sim \text{erfc}\left(\frac{m}{\sqrt{2}}\right)\nonumber\\
 &\quad+ R \Bigg[\left(2 m^2+1\right)
   \text{erfc}\left(\frac{m}{\sqrt{2}}\right)+\left(-4
   m^2-1\right) \text{erfc}\left(\sqrt{2} m\right)-2m
   \sqrt{\frac{2}{\pi }} e^{-2 m^2} \left(e^{\frac{3
   m^2}{2}}-1\right)\Bigg]\,,\quad R\to 0\,,\label{eq:hnoR0}
\end{align}
and for $R\to \infty$ as
\begin{align}
  h_{\text{no bridge}}(R,m)  &\sim R e^{-\sqrt{2} \sqrt{R} m}\,,\quad R\to \infty\,.\label{eq:hnoinf}
\end{align}
Contrary to the expansion of $h(R,m)$ close to $R\!=\!0$ in Eq.~(\ref{eq:hr0a}), the term in brackets in the expansion of $h_{\text{no bridge}}(R,m)$ close to $R\!=\!0$ in Eq.~(\ref{eq:hnoR0}) is not always positive (see red curve in Fig.~\ref{fig:dh}). This implies that the optimal resetting rate for RBM freezes to $0$ beyond a critical value $m=m_c$. The critical value $m_c>0$ is defined such that the term in brackets in Eq.~(\ref{eq:hnoR0}) vanishes:
\begin{align}
  \left(2 m_c^2+1\right)
   \text{erfc}\left(\frac{m_c}{\sqrt{2}}\right)-\left(4
   m_c^2+1\right) \text{erfc}\left(\sqrt{2}\, m_c\right)-2\,m_c
   \sqrt{\frac{2}{\pi }} e^{-2 m_c^2} \left(e^{\frac{3
   m_c^2}{2}}-1\right)=0\,.\label{eq:mc}
\end{align} 
Numerically, we find $m_c\approx 0.820$. In Fig.~\ref{fig:roptz}, we compare the optimal resetting rate $R^*(m)$ that maximizes the probability to reach $m=M/\sqrt{2Dt_f}$ for a RBB and a RBM. 

\subsection{Expected maximum}

In this section, we compute the expected maximum $\langle M(t_f)\rangle$ of a RBB of duration $t_f$. It turns out that this is closely related to the hitting probability discussed
before. The expected maximum is given by
\begin{eqnarray} \label{def_exp}
\langle M(t_f) \rangle = \int_0^\infty M \, \frac{d}{dM} {\rm Prob.}\left[ M(t_f) < M\right]\, dM  \;,
\end{eqnarray}
where $ {\rm Prob.}\left[ M(t_f) < M\right]$ is the cumulative distribution of the maximum of a RBB of duration $t_f$. By noting that
\begin{align}
  \text{Prob.}(M(t_f)<M) = 1 -  p_{\text{hit}}(t_f,M)\,,\label{eq:id}
\end{align}
which states that for the maximum of the RBB to be less that $M$ is must not have hit the level $M$ during the time $t_f$. Inserting Eq.~(\ref{eq:id}) into Eq.~(\ref{def_exp}), we find that the expected maximum is thus given by 
\begin{align}
  \langle M(t_f) \rangle = \int_0^\infty dM p_{\text{hit}}(t_f,M) \,,\label{eq:avgM}
\end{align}
where we used integration by parts.
By inserting the expression of $p_{\text{hit}}(t_f,M)$ given in Eq.~(\ref{eq:phits}) and performing the integration, we find
\begin{align}
   \langle M(t_f) \rangle =\frac{\sqrt{Dt_f}\sqrt{\pi }}{\sqrt{\pi r t_f}\,\text{erf}\left(\sqrt{r t_f}\right)+e^{-r t_f}}\,\int_\Gamma \frac{du\,e^{u}}{2\pi i}\left[ \frac{u }{r t_f^2}\log \left(\frac{u}{r t_f+u}\right) + \frac{1}{r t_f}+ \frac{1}{u}\log \left(\frac{r t_f+u}{u}\right)\right]\,.\label{eq:hRyi}
\end{align}
The remaining Laplace transform can be inverted exactly and gives
\begin{align}
   \langle M(t_f) \rangle = \sqrt{\pi Dt_f}\, q(R=r t_f)\,,\label{eq:Mq}
\end{align}
where the scaling function $q$ is given by
\begin{align}
  q(R) = \frac{ 1-(1+R)e^{-R}+R^2[\gamma+\Gamma(0,R)+\log(R)]}{R^2[e^{-R}+\sqrt{\pi R}\,\text{erf}(\sqrt{R})]}\,,\label{eq:qR}
\end{align}
where $\gamma$ is the Euler-Mascheroni constant and $\Gamma(s,x)=\int_x^\infty dt\, t^{s-1}e^{-t}$ is the incomplete gamma function. The function $q(R)$ has the following asymptotic behavior
\begin{align}
  q(R) &\sim \frac{1}{2} + \frac{R}{6}\,,\quad R\to 0\,,\label{eq:qs}\\
  q(R) &\sim \frac{\log(R)}{\sqrt{\pi R}}\,,\quad R\to \infty\,. \label{eq:qL}
\end{align}
As $q(0)>0$, $\partial_R q(R)\rvert_{R=0}>0$ and $\lim_{R\to\infty}{q(R)}=0$, the function $q(R)$ as a function of $R$ has a non-monotonic behavior and is maximizes at $R^*\approx 2.153$ as can be seen in the right panel in Fig.~3 in the main text. 

%-----------------------------------------------------